\documentclass[pra,letterpaper,twocolumn,showpacs,superscriptaddress,floatfix]{revtex4}
\usepackage{graphicx,psfrag,amsmath,amssymb,amsfonts,bbm,latexsym,color,dcolumn,bm}

\begin{document}

\title{Towards a precision measurement of the Casimir force \\
in a cylinder-plane geometry}

\author{M. Brown-Hayes}

\affiliation{Department of Physics and Astronomy,Dartmouth 
College,6127 Wilder Laboratory,Hanover,NH 03755,USA}

\author{D.A.R. Dalvit}

\affiliation{Theoretical Division,MS B213,Los 
Alamos National Laboratory,Los Alamos,NM 87545,USA}

\author{F.D. Mazzitelli}

\affiliation{Departamento de Fisica J.J. Giambiagi, Facultad de Ciencias Exactas
y Naturales, Universidad de Buenos Aires, 
Ciudad Universitaria, Pabellon 1, 1428 Buenos Aires, Argentina}

\author{W.J. Kim}

\affiliation{Department of Physics and Astronomy,Dartmouth
  College,6127 
Wilder Laboratory,Hanover,NH 03755,USA}

\author{R. Onofrio}

\affiliation{Department of Physics and Astronomy,Dartmouth 
College,6127 Wilder Laboratory,Hanover,NH 03755,USA}

\affiliation{Dipartimento di Fisica ``G. Galilei'',Universit\`a 
di Padova,Via Marzolo 8,Padova 35131,Italy}

\date{\today}

\begin{abstract}

We report on a proposal aimed at measuring the Casimir force in the 
cylinder-plane configuration.
The Casimir force is evaluated including corrections due to finite 
parallelism, conductivity, and temperature. 
The range of validity of the proximity force approximation is also discussed. 
An apparatus to test the feasibility of a precision measurement in
this configuration has been developed, and we describe both a
procedure to control the parallelism and the results of the 
electrostatic calibration.
Finally we discuss the possibility of measuring the thermal
contribution to the Casimir force and deviations from
the proximity force approximation, both of which are expected at 
relatively large distances.

\end{abstract}


\maketitle

\section{Introduction}

The study of quantum vacuum in modern physics is crucial due to
its profound implications over a broad range of lengthscales, from 
elementary particle physics and quantum field theory \cite{Milonni} 
to cosmology \cite{Zeldovich,Weinberg,Carroll, Sahni,Peebles}. 
In between the extremes, the Casimir force \cite{Casimir} 
has provided an experimentally accessible window at the mesoscopic
scale through which significant information 
about quantum vacuum can be retrieved. 
The Casimir force, which can be interpreted 
as the net effect of the radiation pressure resulting from 
the zero point electromagnetic fluctuations, has been studied 
in detail both theoretically and experimentally
\cite{Plunien,Mostepanenko,Bordag,Bordag1,Reynaud,Milton,Milton1,Lamoreauxrev}. 
A first generation of experimental studies immediately followed 
this prediction, both in the parallel plane configuration originally 
proposed by Casimir himself \cite{Sparnaay}, and in a variant of 
this configuration based upon a sphere and a plane \cite{vanBlockland}. 
These attempts had partial success in measuring the Casimir force due 
to a variety of technical issues. In the last decade, a new wave 
of experiments have succeeded in measuring the force in the parallel 
plane and in the sphere-plane geometries. 
The accuracy obtained in the measurements ranges from  15$\%$ in 
the parallel plane case \cite{Bressi} to 0.1-5$\%$ in the sphere-plane 
case \cite{Lamoreaux,Mohideen,Chan,Decca,Decca1}. 
The accuracy for the former configuration is limited mainly by the 
stringent requirements for parallelism between the two plates while  
in the latter configuration the limitation is due to the small 
force signal available, leading to a maximum explorable distance 
between sphere and plane of about 1$\mu$m \cite{Footnote1}. 
At distances smaller than 1$\mu$m the correction to the 
Casimir force due to finite conductivity and roughness of the 
substrates cannot be neglected, and has to be taken into 
account in the theoretical expression of the force. 
Furthermore, the Casimir force in the sphere-plane configuration 
is evaluated by using the so-called proximity force approximation 
\cite{Derjaguin, Blocki}, introducing an uncertainty, estimated 
to be in the 0.1 $\%$ range, in the theoretical prediction.

In this paper we report on theoretical and experimental studies 
of a geometry which interpolates between the two abovementioned
configurations, namely the cylinder-plane geometry. 
This geometry is a compromise between the parallel plane and
sphere-plane configurations, as it offers a simpler way to control 
the parallelism, with respect to the former geometry, while 
providing a sufficiently increased force signal at large distances in 
comparison to the sphere-plane configuration.
 The study of the cylinder-plane configuration also provides insights 
into the finite temperature contribution to the Casimir force, 
as well as into the validity of the proximity force approximation.
We show how some of these open issues in large-distance Casimir
physics, still to be pursued in the laboratory, are easier to deal 
with in this geometry with respect to those already studied.
Mastering the Casimir force at the highest level of accuracy 
is mandatory to give limits to other macroscopic forces acting in the 
micrometer range, such as expected corrections to the Newtonian 
gravitational force \cite{Fishbach}.

The paper is organized as follows: in Section II we discuss the
Casimir force in the cylinder-plane geometry introducing 
the main sources of deviation from ideality such as finite
parallelism, finite conductivity, and finite temperature.
Prior to this, the calculation of the electrostatic force in 
the same geometry is presented. This is crucial not only for the 
calibration of the apparatus, by applying externally controlled 
electric fields, but also for the discussion of the expected background 
noise due to unavoidable residual electrical charges present on 
the two surfaces. Also, the validity of the proximity force 
approximation and some related subtleties in its definition 
at the next-to-leading order are discussed.
In Section III, we present an apparatus developed at Dartmouth
to test the basic principle of the measurement and to demonstrate
various techniques specific to this configuration.
In Section IV, we present the projected sensitivity of the apparatus.
This leads to a discussion of the possible explorable physics, in 
particular the measurement of the thermal contribution to the Casimir 
force and the test of the validity of the proximity force approximation.

\begin{figure}[t]
\includegraphics[width=1.00\columnwidth]{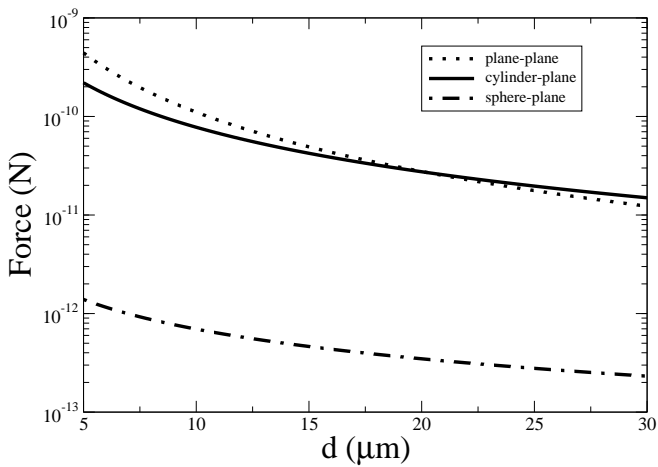}
\caption{Absolute electrostatic forces for parallel plane (dotted curve),
cylinder-plane (solid), and sphere-plane configurations (dot-dashed). 
The force in the plane-plane case is given by $F_{\rm El}^{(0)} =\epsilon_0 A 
V^2/2d^2$, and in the sphere-plane case it is given by $F_{\rm 
El}^{(0)}=2 \pi \epsilon_0 V^2 \sum_{n=1}^{\infty} [ \coth(u) - n 
\coth(n u)]/\sinh(n u)$, with $\cosh(u)=1+d/R$. We use $V=50 {\rm 
mV}$ for the bias voltage, $A=1 {\rm mm}^2$ for the surface area 
of the parallel plane configuration, $L=1 {\rm cm}$ for the 
length of the cylinder, $a=100 \mu{\rm m}$ for the radius of the 
cylinder, and $R=100 \mu{\rm m}$ for the radius of curvature of 
the sphere. For the parallel plane and sphere-plane configurations 
these values are taken from actual experiments described in 
Refs. \cite{Bressi} and \cite{Mohideen}, respectively.} 
\label{fig1}
\end{figure}

\section{Coulomb and Casimir forces in a cylinder-plane geometry}

\subsection{Electrostatic force}

A first step towards measuring the Casimir force in the 
cylinder-plane geometry is to evaluate the force signal 
expected for the corresponding electrostatic force. 
Any apparatus for measuring the Casimir force has to be 
calibrated with a more controllable, better understood, 
force like the Coulomb force. 
Additionally, the electrostatic force is unavoidably 
present as a background due to the residual electric 
charges on the conducting surfaces. Let us consider a perfectly 
conducting cylinder of length $L$ and radius $a$ (with $L \gg a$ 
to neglect border effects)  kept at a fixed electrostatic 
potential $V_0$. The cylinder is parallel to a perfectly 
conducting, grounded, planar surface of area $A$, and the 
distance between the two conductors is denoted by $d$. 
For this geometry, the exact electrostatic force between 
the cylinder and the plane is given by \cite{Smythe}
\begin{equation}
F_{\rm El-ex}^{(0)} = \frac{ 4 \pi \epsilon_0 L V_0^2}{\Delta \ln^2 
\left( \frac{h-\Delta}{h+\Delta} \right)} , \label{forceSmythe}
\end{equation}
where $\Delta=\sqrt{h^2-a^2}$ and $h=d+a$. In the limit $d \ll 
a$, this expression reduces to
\begin{equation}
F_{\rm El}^{(0)} = \frac{\pi \epsilon_0 \sqrt{a} L V_0^2}{2 
\sqrt{2} d^{3/2}}. 
\label{forceELprox}
\end{equation}

\begin{figure}[t]
\includegraphics[width=0.4 \columnwidth]{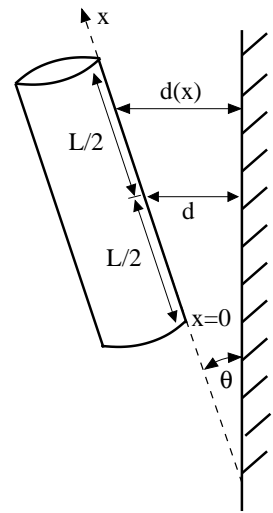}
\caption{Geometrical construction of the non-parallel
configuration. For the proximity approximation,
the local distance between the cylinder and the
plane is $d(x)=d - (L/2 -x) \tan \theta$.}
\label{fig2}
\end{figure}

As we will discuss below, this equation can also be derived using 
the proximity force approximation 
\cite{Derjaguin, Blocki}.
In Fig. \ref{fig1} we compare the absolute electrostatic forces  
corresponding to the parallel plane, cylinder-plane and sphere-plane 
configurations with typical values of the relevant parameters, already 
achieved or achievable in practice. 
Since, under general experimental conditions, the cylinder and 
the plane will not be perfectly parallel, we need to calculate the 
corrections to Eq. (\ref{forceELprox}) due to non-parallelism. 
This is easily obtained by using the proximity force approximation, 
with a local distance between the cylinder and the plane given by 
$d(x)=d - (L/2 -x) \tan \theta$ (see Fig. \ref{fig2}). 
We obtain
\begin{eqnarray}
F_{\rm El}^{\rm np} &=& F_{\rm El}^{(0)}  \frac{1}{\alpha}
\left( \frac{1}{\sqrt{1-\alpha}} - \frac{1}{\sqrt{1+\alpha}} \right) \nonumber \\
&\approx &F_{\rm El}^{(0)} \left[ 1 + \frac{5}{8} \alpha^2 + O(\alpha^4) \right] ,
\label{emnonparallel}
\end{eqnarray}
where $\alpha= L \sin \theta/ 2 d$. As described in the next section,
the quadratic dependence of the electrostatic force on the angle
measuring the deviation from the ideal parallelism provides a way
to optimize the parallelism between the cylinder and the plane 
during the electrostatic calibration.

\begin{figure}[t]
\includegraphics[width=1.00\columnwidth]{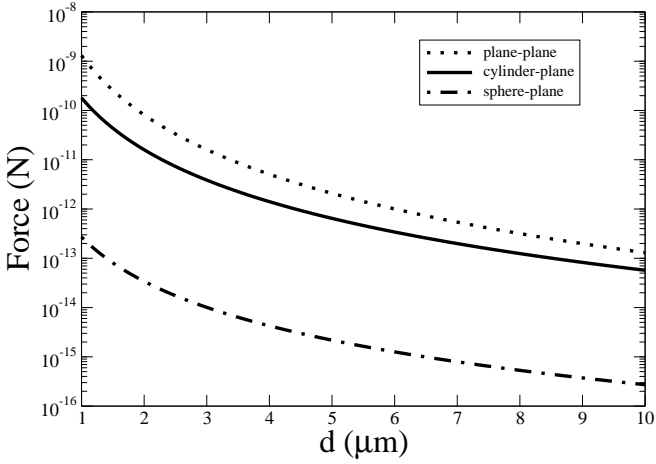}
\caption{Absolute Casimir force for plane-plane (dotted curve),
  cylinder-plane (solid), and sphere-plane configurations
  (dot-dashed).
For the plane-plane case it is given by $F_{\rm Cas}^{(0)} = 
\pi^2 \hbar c A / 240 d^4$. The latter two are evaluated within 
the proximity force approximation. For the sphere-plane case it is
given by $F_{\rm Cas}^{(0)} = \pi^3 \hbar c R / 360 d^3$. 
Parameters are the same as in Fig. \ref{fig1}.}
\label{fig3}
\end{figure}

\subsection{Casimir force}

To evaluate the  Casimir force between the cylinder and the plane 
we use once again  the proximity force approximation. For the 
parallel case ($\alpha=0$), the Casimir force between the 
cylinder and the plane is attractive, and its magnitude in the 
limit $d << a$ is given by \cite{Dalvit}
\begin{equation}
F_{\rm Cas}^{(0)} = \frac{ \pi^3 \hbar c L a^{1/2}}{384 \sqrt{2} 
d^{7/2}}. 
\label{casimircyl}
\end{equation}
The scaling of the Casimir force with distance in the cylinder-plane 
geometry is intermediate between the sphere-plane case ($\propto
d^{-3}$) and the parallel plane configuration ($\propto d^{-4}$). 
The absolute force signal, for typical values of the relevant 
parameters, is also intermediate (see Fig. \ref{fig3}). 
With respect to the sphere-plane geometry, one can enhance 
the signal by exploiting the linear dimension, {\it i.e.} the 
size $L$, as long as the parallelism between the cylinder and 
the planar surface does not become an issue. In comparison 
to the parallel plane situation, in the cylinder-plane configuration 
one needs to parallelize in only one spatial dimension instead of 
two, the latter being a considerably more difficult task.
The correction to the cylinder-plane Casimir force in the slightly 
non-parallel case reads
\begin{eqnarray}
F_{\rm Cas}^{\rm np} &=& F_{\rm Cas}^{(0)} \frac{1}{5 \alpha}
\left( \frac{1}{\sqrt{(1-\alpha)^5}} - \frac{1}{\sqrt{(1+\alpha)^5}}
\right) \nonumber \\
&\approx &F_{\rm Cas}^{(0)} \left[ 1 + \frac{21}{8}
\alpha^2 + O(\alpha^4) \right],
\end{eqnarray}
which shows a strong similarity to the non-parallel Coulomb force, 
just differing at the leading orders by the coefficient of the 
quadratic correction in the parameter $\alpha$.
  
For an accurate comparison between experiment and theory, apart from
the deviations from parallelism already taken into account within the
proximity force approximation scheme, we consider the deviations
of the predicted force from the ideal situation of perfect conductors,
zero roughness, and zero temperature. For typical surfaces and realistic
experimental sensitivities, the roughness correction is negligible
with respect to other deviations at the
distances we are interested in ($d > 1 \mu{\rm m}$).
\begin{figure}[t]
\includegraphics[width=1.00\columnwidth]{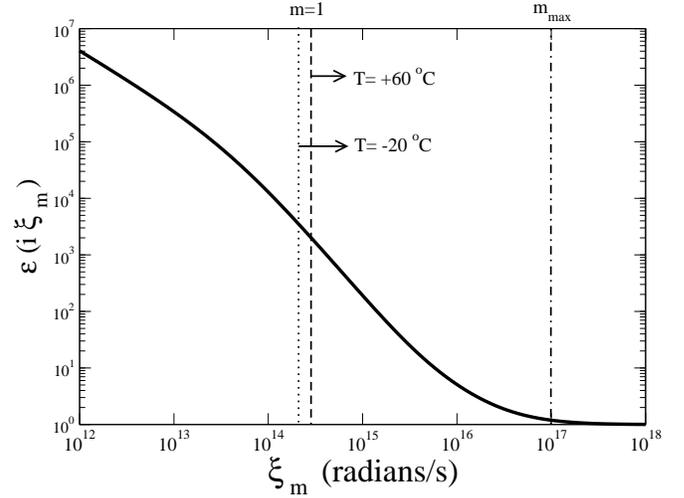}
\caption{Permittivity for Au as a function of frequency
calculated from optical data (courtesy of Astrid Lambrecht and Serge Reynaud).
Permittivity data for Matsubara frequencies with $m \ge1$ in the range
of temperatures of interest $-20 ^o C \le T \le +60 ^oC$ can be
obtained from this plot. The vertical lines show the frequency region 
used in the $m$ summation over the Matsubara frequencies. The dotted line
corresponds to $m=1$ at $T=-20 ^oC$, the dashed line to $m=1$ at
$T=+60 ^oC$, and the dotted-dashed line to the maximum value $m_{\rm
max}$ used at all temperatures, corresponding to a maximum Matsubara 
frequency of $\xi_{\rm max}=10^{17} {\rm rad/s}$.}
\label{fig4}
\end{figure} 
On the other hand, combined temperature and conductivity corrections
are usually important in this range of distances.  We have computed these
corrections via the Lifshitz formalism \cite{Lifshitz}, which provides an
expression for the pressure between two infinite, parallel
plates. We have then used this result in the proximity force
approximation for the cylinder-plane configuration. 
The Casimir pressure in the plane-plane configuration at finite 
temperature $T$ is given by the Lifshitz formula
\begin{eqnarray}
P(d) &=& - \frac{1}{\pi \beta d^3} \sum_{m=0}^{\infty \; '} \int_{m \gamma}^{\infty} dy \; y^2
\times \nonumber \\
&& \left[  \frac{r_{\rm TM}^{-2} e^{-2y}}{1-r_{\rm TM}^{-2} e^{-2y}}  +
\frac{r_{\rm TE}^{-2} e^{-2y}}{1-r_{\rm TE}^{-2} e^{-2y}} \right] ,
\label{pressurepp}
\end{eqnarray}
where $d$ is the gap between the plates, $\beta=1/k_{\rm B} T$ is the
inverse temperature, and $\gamma=2 \pi d / \beta \hbar c$. 
The prime on the summation sign indicates that the $m=0$ term 
is counted with half weight. The reflection coefficients
$r_{\rm TE}$ and $r_{\rm TM}$ for the two independent polarizations 
TE and TM are computed at imaginary frequencies $\omega_m=i \xi_m$, 
where $\xi_m=2 \pi m / \beta \hbar$ are the Matsubara frequencies.

\begin{figure}[t]
\includegraphics[width=1.00\columnwidth]{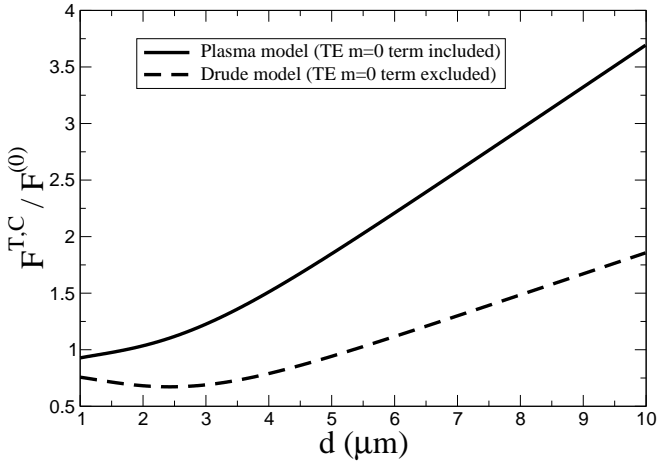}
\caption{Combined thermal and conductivity corrections to the Casimir
force in the cylinder-plane geometry. We show the Casimir force, 
normalized to its bare value for infinite conductivity and 
zero temperature, versus the distance $d$ in the case of gold 
metallic surfaces at a temperature $T=300 ^o {\rm K}$.
Optical data are used to calculate the frequency-dependent 
permittivity of Au. The $m=0$ contribution to the Lifshitz formula 
is computed by extrapolating the optical data to zero frequency 
using two different theoretical approaches:  the plasma model 
(for which the TE $m=0$ mode contributes to the force), and  
the Drude model (for which the TE $m=0$ mode does not contribute).
Parameters are the same as in Fig. \ref{fig1}.}
\label{fig5}
\end{figure}

Although the foundations for the Lifshitz formula are well
established, the exact expressions for the reflectivity 
coefficients are not. Following the Lifshitz formalism, the 
reflection coefficients are expressed in terms of the dielectric 
permittivity $\epsilon(\omega)$ as
\begin{eqnarray}
r_{\rm TM}^{-2} = \left[ \frac{\epsilon(i \xi_m) p_m + s_m}{\epsilon(i \xi_m) p_m - s_m} \right]^2  &;&
r_{\rm TE}^{-2} =\left[ \frac{s_m + p_m}{s_m - p_m} \right]^2 ,
\end{eqnarray}
where $p_m=y/ m \gamma$ and $s_m = \sqrt{\epsilon(i  \xi_m) - 1 + p_m^2}$.

\begin{figure}[t]
\includegraphics[width=1.00\columnwidth]{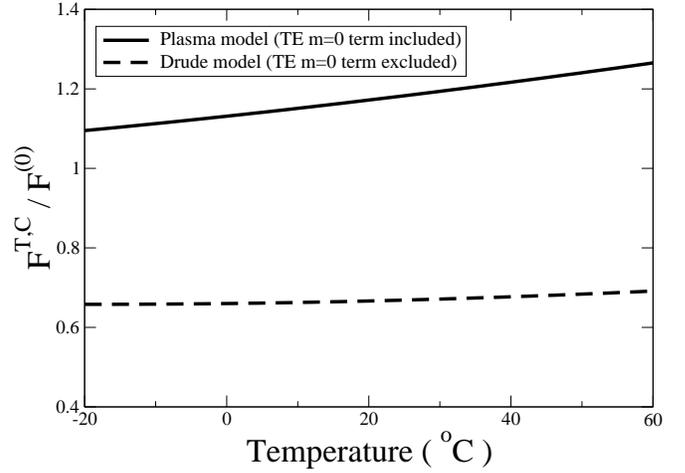}
\caption{Dependence upon temperature of the Casimir
force in the cylinder-plane geometry. We show the Casimir force,
normalized to its bare value for infinite conductivity and zero
temperature, versus the temperature $T$ in the case of gold metallic
surfaces separated by a distance $d=3 \mu{\rm m}$.
Extrapolated data for zero frequency ($m=0$) is obtained from two different
theoretical approaches: the plasma model (for which the TE $m=0$
reflection coefficient is non-zero), and the Drude model (for which 
the TE $m=0$ reflection coefficient vanishes).
Parameters are the same as in Fig. \ref{fig1}.}
\label{fig6}
\end{figure}

Using tabulated optical data for different metals \cite{Palik}, it is 
possible to compute the dielectric permittivity along the imaginary 
frequency axis \cite{Lambrecht}.  As an example, we show in 
Fig. \ref{fig4} the numerically computed permittivity of Au as a 
function of frequency. 
For the computation of the $m$ summation in Eq. (\ref{pressurepp}) we
used a cut-off $m_{\rm max}$ corresponding to a Matsubara frequency
$\xi_{m_{\rm max}}=10^{17} {\rm rad/sec}$.  
For the range of temperatures we are interested in ($-20 ^o C \le T
\le + 60 ^o C$), permittivity data for all Matsubara frequencies 
$\xi_m=2 \pi m / \beta \hbar$ corresponding to $m \ge 1$ can
be extracted from the optical data (see Fig. \ref{fig4}).
To calculate the $m=0$ contribution, it is however necessary to extrapolate
the available data to zero frequency. This extrapolation has been 
done in the literature using different theoretical models, and has 
led to controversial predictions for the Casimir force between parallel plates
\cite{Bostrom,Lamoreaux1,Chen,Geyer,Esquivel,Genet,Svetovoy,Decca1,Torgerson,Esquivel1,Hoye,Bordag2}.

We have computed the Casimir force between the cylinder and the plane 
using two distinct theoretical approaches. 
In the first approach, the optical data are extrapolated 
using the plasma model for the dielectric permittivity \cite{Bordag1,Decca1}: 
$\epsilon(i \xi) = 1 + \omega_p^2/\xi^2$, 
where $\omega_p$ is the plasma frequency (equal to 9.0 eV for Au). In
this model, the reflectivity coefficients for $m=0$ are given by
\begin{eqnarray}
r_{\rm TM}^{-2}(m=0) &=&1 , \nonumber \\
r_{\rm TE}^{-2}(m=0) &=& \left[
\frac{c y/d + \sqrt{\omega_p^2 + (c y/d)^2 }}{c y/d - \sqrt{\omega_p^2 + (c y/d)^2}}
\right]^2 .
\label{reflectivity_plasma}
\end{eqnarray}
\begin{figure}[t]
\includegraphics[width=0.4\columnwidth]{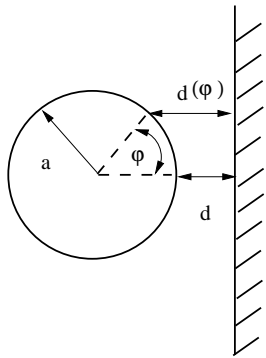}
\caption{Cylinder-plane geometry (lateral view). 
In the proximity force approximation, the force 
is computed as a superposition of forces between infinitesimal 
parallel plates, separated by a distance $d(\varphi)=d+a(1-\cos\varphi)$.}
\label{fig7}
\end{figure}

In the second approach, we use the model of \cite{Hoye}, that
extrapolates the optical data using the Drude model:  
$\epsilon(i \xi) = 1 + \omega_p^2/ \xi (\xi + \nu)$, where $\nu$ 
is the relaxation frequency (equal to 35 meV for Au). 
In this second model, the reflectivity coefficients for $m=0$
are, in contrast to those found in Eq. (\ref{reflectivity_plasma}),  
\begin{eqnarray}
r_{\rm TM}^{-2}(m=0) = 1 &;& r_{\rm TE}^{-2}(m=0) = 0 .
\end{eqnarray}
That is, in this second approach the transverse electric zero 
mode does not contribute to the Casimir force.
In Fig. \ref{fig5} we  show the ratio $F_{\rm Cas}^{T,C} / F^{(0)}$ 
between the real Casimir force (including temperature and 
conductivity corrections) and the ideal one (perfect conductors, 
zero temperature) as a function of the gap $d$ between the cylinder and the plane, assumed
to be parallel. The two curves correspond to the two different theoretical approaches
described above. As follows from the figure, the ratio of forces
increases monotonically with distance for the plasma model, while it
shows a small dip close to $3 \mu{\rm m}$ for the Drude model. 
In Fig. \ref{fig6} we show the same ratio of forces as a function of
temperature for a fixed gap between the cylinder and the plane, set 
at $d=3 \mu{\rm m}$. The temperature range corresponds to the one 
we expect to control during the actual Casimir experiment at finite
temperature.  We see that the normalized force increases with
temperature much faster for the plasma model than for the Drude model.
In the entire targeted range of temperatures the two models give 
predicted forces differing by roughly a factor of two, allowing for 
an easier experimental discrimination, provided that thermal expansion
of the materials of the experimental set-up will be kept under control.

\subsection{Accuracy of the proximity force approximation}

The calculations of the electrostatic and Casimir forces done in 
the previous sections rely on the proximity force approximation. 
We now discuss its validity in the ideal case of zero temperature 
and perfect conductor. 

Let us first consider the proximity force approximation to the 
electrostatic force in the parallel cylinder-plane configuration
\begin{equation}
F^{(i)}_{\rm El}=\frac{\epsilon_0V_0^2}{a^2}\int_0^{\pi/2}
\frac{dA_i}{(1+\frac{d}{a}- \cos\varphi)^2}\,\,\, 
. \label{var-el}
\end{equation}
Here $dA_i$ is the effective area of the infinitesimal parts in 
which the surfaces are divided to integrate the parallel plates 
result, and $\varphi$ is the angle parameterizing the location 
of the infinitesimal surfaces on the the cylinder (see Fig. \ref{fig7}).
Different choices for this area give distinct proximity 
approximations for the force \cite{Footnote2}. For example, the area of a small 
portion of cylinder is $dA_c=Lad\varphi$, while the area of a 
small portion of plane is $dA_p=La\cos\varphi d\varphi$. 
One could also use a combination of the two, like the geometric 
mean $dA_{gm}=(dA_pdA_c)^{1/2}$. These choices for the effective 
area give the same result to leading order in $d/a$, but they 
differ in the subleading corrections
\begin{equation}  
\frac{F^{(i)}_{\rm El}}{F^{(0)}_{\rm El}} = 1+\eta^{(i)}_{\rm El}  
\frac{d}{a} + 
O\left( \frac{d^2}{a^2} \right)\,\,\, ,
\label{var-el-result}
\end{equation}
where $\eta^{(p)}_{\rm El}=-0.75$, $\,\, \eta^{(c)}_{\rm El}=0.25$,
and $\eta^{(gm)}_{\rm El}=-0.25$ \cite{Footnote3}. 

One could estimate the error of the proximity force approximation 
as the difference between these results. However, for the 
parallel cylinder-plane configuration we know  the exact value of 
the electrostatic force, so we can use it to analyze the accuracy 
of the different versions of the proximity force approximation. 
The ratio between the exact and the leading proximity 
electrostatic forces Eqs. (\ref{forceSmythe}) and 
(\ref{forceELprox}) is given by 
\begin{equation}
\frac{F^{(0)}_{\rm El-ex}}{F^{(0)}_{\rm El}}= 1-0.083 \frac{d}{a} + 
0.035   \frac{d^2}{a^2}  + \ldots \,\,\, . \label{pfa-ex}
\end{equation}
Both results coincide within $1\%$ for $d/a<0.12$. 

\begin{figure}[t]
\includegraphics[width=1.00\columnwidth]{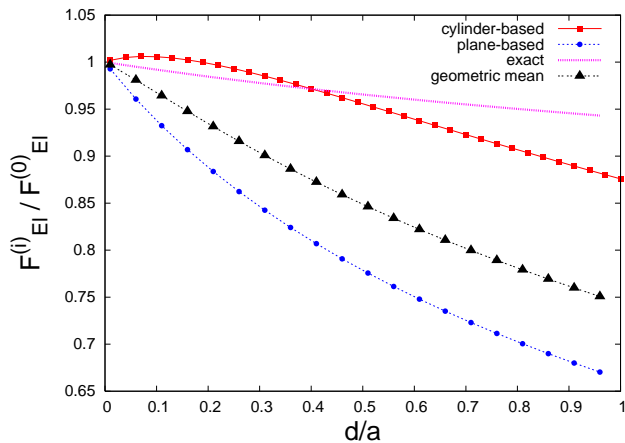}
\caption{(Color online) Electrostatic force in the cylinder-plane 
geometry, normalized to the electrostatic force evaluated in the 
leading proximity approximation of Eq. (\ref{forceELprox}).
We show the exact force (continuous line), and the force evaluated 
through different proximity approximation schemes obtained using the
area of the cylinder (squares), the area of the plane (dots), and 
a geometric mean of the areas (triangles).}
\label{fig8}
\end{figure}

In Fig. \ref{fig8} we show the ratios
$F^{(i)}_{\rm El}/F^{(0)}_{\rm El}$, together with
the exact result, also normalized to the leading proximity force 
approximation in Eq. (2). 
For small $d/a$, the  exact value of the force  is in between 
$F^{(c)}_{\rm El}$ and $F^{(gm)}_{\rm El}$.
Moreover, a comparison of Eqs. (\ref{var-el-result}) and
(\ref{pfa-ex}) shows that the geometric mean prescription is closer 
to the exact force (a similar result has been obtained for the Casimir 
interaction energy between concentric cylinders \cite{Stecher}). 
Had we estimated the error of the proximity force approximation using 
$F^{(c)}_{\rm El}$ and $F^{(gm)}_{\rm El}$, we would have concluded 
that it is smaller than $1\%$ for $d/a<0.04$. 

\begin{figure}[t]
\includegraphics[width=1.00\columnwidth]{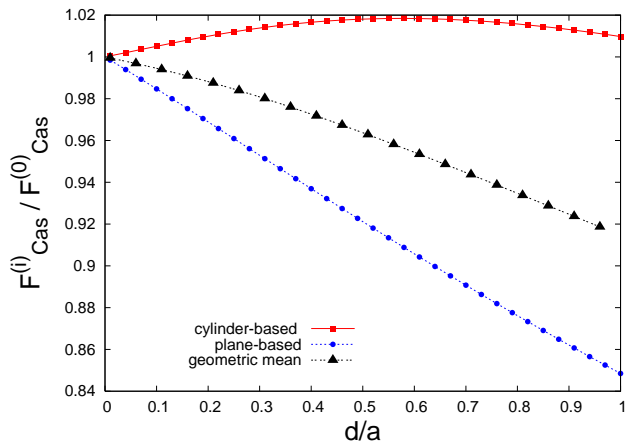}
\caption{(Color online) Casimir force in the cylinder-plane geometry, 
normalized to the Casimir force evaluated in the leading proximity 
approximation. We show the different proximity approximation schemes 
obtained using the area of the cylinder (squares), the area of the
plane (dots), and a geometric mean of the areas (triangles).}
\label{fig9}
\end{figure}

For the Casimir case, an analytic expression for the exact force 
is not available.  Therefore, we will estimate the error of the 
proximity force approximation using 
$F^{(c)}_{\rm Cas}$ and $F^{(gm)}_{\rm Cas}$, defined as 
\begin{equation}
F^{(i)}_{\rm Cas}=\frac{\pi^2\hbar c}{120 a^4}
\int_0^{\pi/2}\frac{dA_i}{(1+\frac{d}{a}- \cos\varphi)^4}\,\,\,  . 
\end{equation}
We obtain 
\begin{equation}  
\frac{F^{(i)}_{\rm Cas}}{F^{(0)}_{\rm Cas}} = 1+\eta^{(i)}_{\rm Cas}
\frac{d}{a} + O\left( \frac{d^2}{a^2} \right) , 
\label{var-cas-results}
\end{equation}
where $\eta^{(p)}_{\rm Cas}= -0.15$,$\,\, \eta^{(c)}_{\rm Cas}= 0.05$
and $\eta^{(gm)}_{\rm Cas}= -0.05$. These results are shown
in Fig. \ref{fig9}. We see that, for small $d/a$, the cylinder-based
proximity approximation is larger than the leading order, while 
the opposite happens for the results based on the plane and 
on the geometric mean areas. This is analogous to the electrostatic case.
Assuming that the exact result is in between $F^{(c)}_{\rm Cas}$ 
and $F^{(gm)}_{\rm Cas}$, we estimate the error of the proximity 
force as smaller than $1\%$ for $d/a <0.2$.
It is worth to note that the spread of the different approximations 
is smaller for the Casimir force than for the electrostatic force 
(see Figs. \ref{fig8} and \ref{fig9}). This is due to the fact that, as 
the Casimir force is stronger than the electrostatic one at small
distances, the proximity force approximation is dominated by a smaller 
region around $\varphi=0$, where the difference between effective 
area is less important.

The exact Casimir energy for massless scalar fields satisfying Dirichlet 
boundary conditions in the cylinder-plate geometry has been 
computed using numerical simulations based on the wordline 
approach to quantum field theory \cite{Gies}. Based on an 
analytic fit of the numerical data \cite{Gies1}, we have found a 
deviation in the force with respect to the leading proximity 
force of $1\%$  already at $d/a=0.06$. The difference is about  
$5\%$ at $d/a=0.5$. These are still preliminary results, because 
the precision of the data is not very high,  on the order of 
$10\%$ for small values of $d/a$ \cite{Gies1}.

\begin{figure*}[t]
\includegraphics[width=1.0\textwidth]{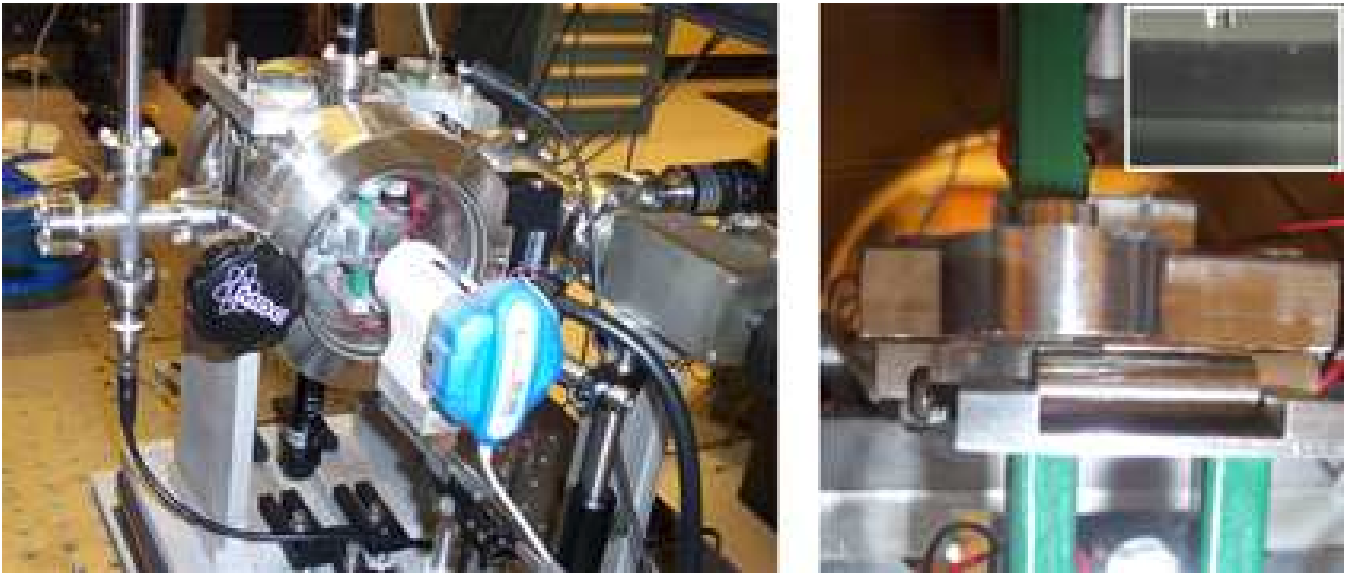}
\caption{(Color online) Images of the experimental set-up, with an overall 
view (left), a close-up on the cylinder-resonator region (right), 
and an image from the optical microscope (inset). 
Inside the vacuum chamber are visible the two piezoelectric actuators 
for the fine approach between the cylinder and the resonator on the 
bottom side, and the piezoelectric actuator for the fine approach 
of the optical fiber and its mount on the top side, with the fiber 
end facing the middle point of the resonator. 
In between are the resonator and the cylinder, with coarse motion 
controlled by vertical feedthrough micrometers.  
The fiber is sent through a feedthrough inside the vacuum chamber 
(left side in the overall view), while a goniometer stage is 
used for coarse parallelization (right side in the overall view).  
The large viewport allows for use of a digital optical microscope with
up to 100 $\times$ magnification for a rough parallelization and a coarse 
assessment of the fiber-resonator and resonator-cylinder distances, as 
visible in the inset.
The size of the resonator is 2 cm $\times$ 1 cm $\times$ 
178 $\mu$m for width, length, and thickness, respectively. 
The cylinder has a diameter $2a=6.35$ mm (1/4 inch) and a length of 2 cm.}  
\label{fig10}
\end{figure*}

Summarizing, if the Casimir force is measured at the $1\%$ 
level,  in order to see deviations from the proximity force the 
ratio $d/a$ should be larger than $0.2$, according to the 
estimation based on the different choices of the area. 
On the other hand, numerical simulations suggest that deviations 
can already be present at $d/a=0.06$. 
Higher precision numerical data for the electromagnetic field 
are needed to confirm this prediction.
 
\section{Electrostatic calibrations}

In order to assess the sensitivity and the requirements for a
precision measurement of the Casimir force in a cylinder-plane  
configuration, we have performed electrostatic calibrations with a 
prototype of the experimental apparatus (see Fig. \ref{fig10} for details). 
The core part of the system is a stainless steel cantilever resonator
faced on opposite sides by a stainless steel cylinder and an optical fiber 
for the detection of its displacement. 
The resonator is clamped to an aluminum base which is in thermal 
contact with a thermoelectric cooler for temperature stabilization.
Below the resonator is the cylinder, attached to a frame 
mounted on two piezoelectric actuators providing a maximum displacement of 
$(15.00 \pm 1.5) \mu$m for an applied voltage of 100 V. 
The two piezoelectric actuators are located along the axis of 
the cylinder, providing both necessary adjustments for parallelization 
and the fine approach of the cylinder to the resonator. 
The coarse approach is provided  by a feedthrough micrometer located at the bottom of the vacuum chamber. 
A few tens of micrometers above the resonator is the optical
fiber, which is part of a fiber optic interferometer \cite{Rugar} 
having as light source a 5 mW diode laser at a wavelength of 671 nm. 
The displacement signal results from the interference between the
light reflected by the resonator and the light internally reflected 
by the fiber end.  
\begin{figure}[t]
\includegraphics[width=1.00\columnwidth]{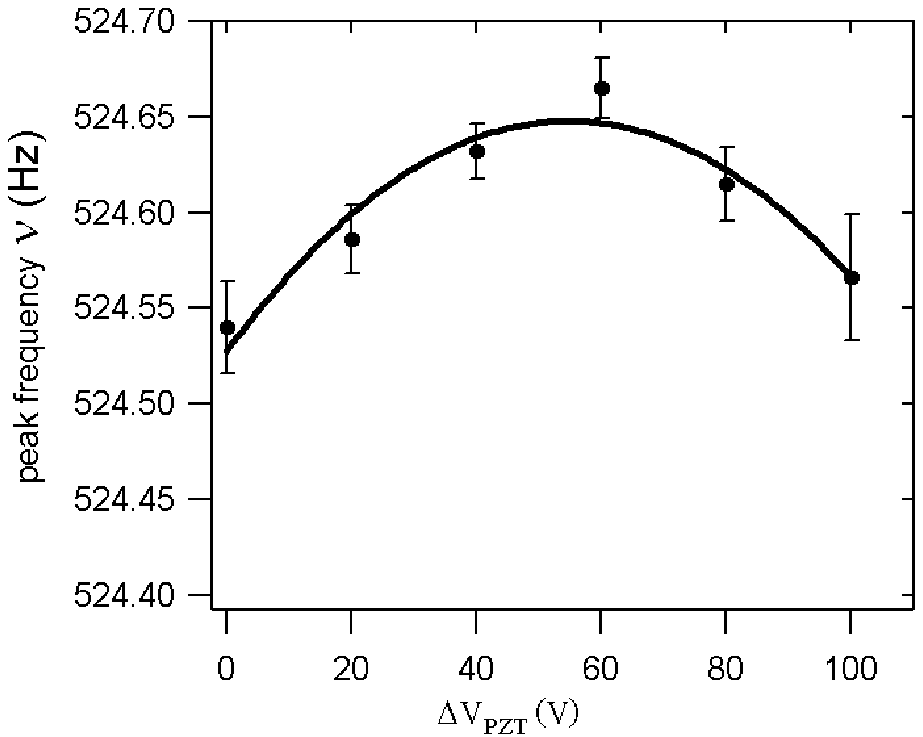}
\caption{Assessment of the parallelism. 
Resonator frequency versus the difference between the voltages 
applied to the two piezoelectric actuators driving the 
cylinder base $\Delta V_{\rm PZT}=V_L-V_R$, starting 
from a common mode of $V_L=V_R=$50 V, for $V_0=100$ V. 
The resonator frequency without bias voltage is $\nu_0=(527.30 \pm 0.02)$ Hz.  
The maximum frequency (corresponding to the minimum frequency shift
with respect to $\nu_0$) realizes the condition for parallelism, 
and this occurs at $\Delta V_{\rm PZT}=(54.8 \pm 7.1)$ V. 
As the two piezoelectric actuators are spaced by 2 cm, this
corresponds to achieving parallelism within 
$\delta \theta=5.3 \times 10^{-5}$ radians. 
The determination of the frequency is based upon a fit of the
mechanical transfer function of the resonator with a Lorentzian 
function, obtained through 100 averages of the FFT with a 
frequency span of 12.5 Hz.}
\label{fig11}
\end{figure}
The condition of interference which maximizes the displacement 
sensitivity (obtained at distances for which the dependence of the 
intensity versus the distance has the maximum slope) is obtained 
with a coarse approach of the optical fiber to the resonator 
through a micrometer stage, and a finer tuning with a piezoelectric
actuator. The fiber position is stabilized by a servo loop circuit 
controlling  the voltage driving the piezoelectric actuator. 
The explored resonator-cylinder distances, ranging between 
15 $\mu$m and 40 $\mu$m, are estimated with a digital microscope,
allowing for a consistency check with the {\it a posteriori} 
determination of the gap through data fitting, as described below.
Further details on the apparatus setup, its characterization,
sensitivity and the noise limitations will be provided in a future publication. 

The calibration has been performed by measuring the frequency shift
induced by the electrostatic force on the resonator (see for instance \cite{Bressi2}). 
For a generic distance-dependent force (such as the Coulomb 
force in Eq. (1) or the Casimir force in Eq. (4)), the shift in the 
proper frequency of the resonator can be written as
\begin{equation}
\Delta \nu^2= \nu^2 -\nu_0^2=-\frac{1}{4 \pi^2 m} \frac{\partial
  F(d)}{\partial  d},
\end{equation}
where $m$ is the effective mass of the mode of oscillation of resonator. 
The corresponding frequency shift for the Coulomb case  then assumes the form
\begin{equation}
\Delta \nu^2_{\rm El}=- \frac{3 \epsilon_0}{16 \sqrt{2} \pi} 
\frac{\sqrt{a} L V_0^2}{m d^{5/2}}=k_c V_0^2,
\end{equation}
where we have introduced a {\sl curvature} parameter $k_c$ to 
parameterize the parabolic behavior of $\Delta \nu^2_{\rm El}$ 
versus the applied bias voltage $V_0$. The determination of the resonant 
frequency of a mode of oscillation of the cantilever is obtained 
by driving its motion with a piezoelectric actuator clamped 
in the proximity of its base and fed by the white noise source 
of a FFT spectrum analyzer. This last is used to acquire and 
perform the spectral analysis of the signal coming from the 
photodiode collecting the interference light at one port of the fiber mixer.

The electrostatic calibrations can be divided into three steps: 
a) determination of the parallel configuration by looking at  
the minimum frequency shift at constant average distance and 
various tilting angles between the resonator and the cylinder, 
b) measurements of the frequency shift versus bias voltage 
in the parallel configuration, and c) repetition of the previous  
measurements for various values of the cylinder-resonator distance.

In order to determine the parallel configuration, we use the result 
obtained in Eq. (\ref{emnonparallel}), according to which the force 
exerted on the cantilever is expected to have a parabolic dependence 
on the angle describing the deviation from the ideal parallelism for 
small angles $\theta$. 
The coarse control of the parallelism is obtained by using a goniometer 
stage on which the resonator is mounted, which can be manually 
controlled with an in-vacuum feedthrough. For the fine control, 
two piezoelectric actuators are used in a differential mode, 
in such a way that the median distance between the resonator 
and the cylinder (i.e. the distance $d$ between the midpoints 
of the two structures, see Fig. \ref{fig2}) is unchanged. 
This is obtained by pivoting the cylinder around the middle point,  
{\it i.e.} by summing and subtracting equal amounts of voltage supplied 
to the two piezoelectric actuators moving the cylinder position. 
The plot of the resonator frequency versus the difference between 
the voltages applied to the left and right piezoelectric actuators 
acting on the cylinder, for a constant bias voltage difference, has 
a parabolic dependence for small deviations from parallelism, with 
its maximum corresponding to the parallel case $\alpha=0$.
\begin{figure}[t]
\includegraphics[width=1.00\columnwidth]{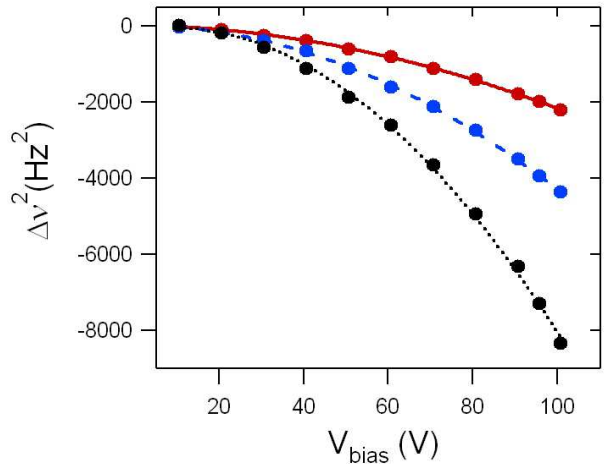}
\caption{(Color online) Electrostatic calibrations. 
Squared frequency shift $\Delta \nu^2$ versus bias voltage 
between the cylinder and the resonator for different values 
of the cylinder-plane distance, corresponding to the 
application of $V_{\rm PZT}=$0 V (continuous line), 
30 V (dashed line), and 70 V (dotted line). 
The curves are the best fit with a generic parabola. 
The typical resonance frequency without bias potential is around 527
Hz, with a bandwidth of 4 Hz. The peak frequency is determined through 
the Lorenzian fit with a typical error of 12 mHz.}
\label{fig12}
\end{figure}
The outcome of this procedure is summarized in Fig. \ref{fig11}, where the
frequency of the resonator displays a parabolic dependence
as a function of $\Delta V_{\rm PZT}$, which reflects 
different tilting angles $\theta$.  
 The maximum frequency defines the parallelism condition, 
within the error, which leads to a precision of  
$\delta \theta= 5.3 \times 10^{-5}$ radians. 
This corresponds, based on Eq. (5), to a correction equal to 
$(F_{\rm Cas}^{\rm np}-F_{\rm Cas}^{(0)})/F_{\rm Cas}^{(0)}=
1.43 \times 10^{-4}$.  
Notice that, unlike the parallel plane configuration, the search for 
the parallel situation is considerably simpler and faster to implement
in the cylinder-plane geometry as it requires only a one-dimensional 
optimization \cite{Footnote4}.
 
\begin{figure}[t]
\includegraphics[width=1.00\columnwidth]{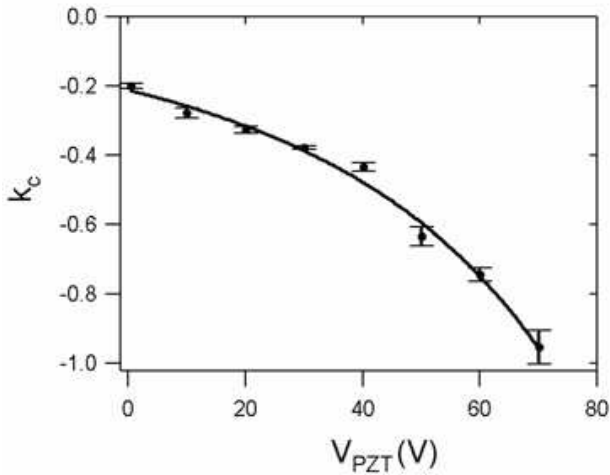}
\caption{Electrostatic calibrations. Plot of the curvature coefficient 
$k_c$ versus the piezoelectric common voltage $V_{\rm PZT}$ and
best fit with a law $\Delta \nu^2=a+b/(V_{\rm MAX} - 
V_{\rm PZT})^{2.5}$ corresponding to the expected frequency-shift 
dependence from Eq. (16). The best fit gives a parameter 
$V_{\rm MAX}=(177.4 \pm 24.1)$V, corresponding to the voltage 
which should close the gap between the cylinder and the resonator.}
\label{fig13}
\end{figure}

After this preparatory measurement, we have then obtained the
frequency shift versus the bias voltage, and the related curvature 
parameter $k_c$ of the expected parabolic dependence, as 
shown in Fig. \ref{fig12}. 
This has been repeated for various distances between the resonator and 
the cylinder, adding a common mode voltage to the actuators, thereby  
inducing a global approach of the cylinder to the resonator. 
In Fig. \ref{fig13} the curvature parameter $k_c$ is plotted versus the 
piezoelectric actuator voltage $V_{\rm PZT}$.  
The latter quantity is related to the absolute distance between the cylinder 
and the resonator as  
$d=d_{\rm in}-\alpha_{\rm act} V_{\rm PZT}$, where 
$\alpha_{\rm act}=(150.0 \pm 15.0) $nm/V is the actuation 
coefficient of the piezoelectric and $d_{\rm in}$ the initial 
distance corresponding to $V_{\rm PZT}=0$.
This allows comparison of the data to the predictions from Eq. (16) 
of a power-law dependence with scaling $-5/2$ upon the absolute distance. 
As usually done in this type of measurement, the absolute initial distance has 
been determined by considering an offset as a free parameter in the fit. 
A zero distance, shorting the cylinder-plane gap, is reached when the 
piezoelectric actuator voltage is 
$V_{\rm PZT}=d_{\rm in}/\alpha_{\rm act} = V_{\rm
  MAX}$. From the value of $V_{\rm MAX}$ determined from the fit, 
$V_{\rm MAX}=(177.4 \pm 24.1)$V, we deduce an initial distance 
$d_{\rm in}=\alpha_{\rm act} V_{\rm MAX}= (26.6 \pm 4.5) \mu$m. 
The estimated minimum gap corresponding to the maximum excursion of
the piezoelectric actuator before shorting the gap is therefore 
$\simeq 16.1 \mu$m, consistent with the expected roughness of 
both cylinder and resonator surfaces. 
The relative accuracy in the determination of the absolute 
distance in these preliminary calibrations is $17 \%$.
Better precision can be obtained by looking at gaps in the 10-30
$\mu$m range using a large magnification, high resolution optical
microscope. The absolute gap can be accurately determined due to 
a particular feature of the cylinder-plane configuration: 
by properly illuminating the gap, the cylinder will appear reflected on 
the planar surface, and the distance between the cylinder edge and
its mirror image on the plane surface can then be measured. 
Extension of the calibration into the range of interest for the 
measurement of the Casimir force (1-5 $\mu$m) will require 
high precision nanopositioners with minimal hysteresis, thereby
reducing the relative error in the determination of the distance to 
less than 1$\%$.

\section{Expected Casimir force signal}

Based on the electrostatic calibrations and the possible short-term 
improvements, we now evaluate the sensitivity to physics related 
to the Casimir effect, in particular the thermal contribution and 
the test of the validity of the proximity force approximation scheme.
The expected frequency shift due to the Casimir force is written as
\begin{equation}
\Delta \nu^2_{\rm Cas-cp}=- \frac{7 \pi}{3072 \sqrt{2}} \frac{\hbar c
  L \sqrt{a}}{m d^{9/2}}.
\end{equation}
It is worth comparing the expected frequency-shift signal to the 
one already measured in Padova \cite{Bressi}. 
The squared frequency shift for a parallel plane configuration is expressed as

\begin{equation}
\Delta \nu^2_{\rm Cas-pp}= -\frac{\hbar c}{240} \frac{A}{m d^5},
\end{equation}
where $A$ is the surface area of the plates.
Considering resonators with the same mass for the two configurations 
corresponding to Eqs. (19) and (20), we obtain a ratio between the 
expected squared frequency shifts as:

\begin{equation}
\frac{\Delta \nu^2_{\rm Cas-cp}}
     {\Delta \nu^2_{\rm Cas-pp}}= 
\frac{35 \pi}{64 \sqrt{2}} \frac{L \sqrt{a d}}{A}.
\end{equation}
Apart from a numerical factor of order unity, the ratio between 
the expected squared frequency shifts is the ratio between 
the relevant geometrical scales in the two configurations, 
{\it i.e.} the transversal size of the resonator-cylinder region $L$, 
the geometrical average of the radius of curvature of the cylinder 
$a$ and the cylinder-plane distance $d$, and the plates 
surface area $A$. By inserting the respective values from the 
Padova experiment, and a radius of the cylinder of 51 cm 
(using commercially available cylindrical lenses with proper 
metallic coating), at the distance $d=1 \mu$m the ratio 
gives $\Delta \nu^2_{\rm Cas-cp}/\Delta \nu^2_{\rm Cas-pp} \simeq 0.727$. 
This implies that a frequency shift equal to 
$\Delta \nu_{cp}=\Delta \nu^2_{cp}/2 \nu_0=$4.08 mHz (571 mHz) at 
a gap of $3 \mu$m (1$\mu$m) is expected for the cylinder-plane case, 
as compared to  a $\Delta \nu_{pp}=$3.2 mHz (780 mHz) at a gap of 
$3 \mu$m (1$\mu$m) for the parallel plane case. 
In preliminary long time average tests with our prototype we have 
obtained an error on the determination of the peak frequency of 
the resonator through a Lorenzian fit of $\simeq 6$ mHz. 
We find that the minimum detectable frequency shift 
can be made significantly smaller 
through a careful study of the resonance curve of the mechanical 
transfer function of the cantilever in a relatively large range 
of frequencies, equal to 10-20 times its intrinsic bandwidth. 
A careful fitting of data, with a long sampling time and a large 
number of averages, allow for the determination of the resonance 
frequency with a precision of the order of  mHz or less, 
even when  working with relatively large mechanical bandwidths as 
in the case of stainless steel resonators. From this point of 
view the use of resonators with a large quality factor may not seem 
advantageous; in order to acquire a resonance curve with the same resolution
one needs to adapt a smaller frequency binning resulting in a much 
longer integration time, which is more vulnerable to 
frequency shifts possibly induced by the finite degree of
stabilization of the temperature of the apparatus.  
Further improvements in the sensitivity are expected by implementing 
frequency and amplitude stabilization of the diode laser,
high-performance actuators with minimal hysteresis and better 
displacement resolution, and a better control of the temperature 
setting and stabilization both for the resonator and for the fiber set-up. 
Alternative detection schemes, like homodyne or heterodyne modulations 
\cite{Bressi2}, may be utilized to improve the precision. 

As a final assessment of the experimental set-up, apart from the need 
for improved planarity and decreased roughness of the resonator and 
cylinder surfaces using optical quality surfaces, and a careful determination of the absolute 
distance between the cylinder and the resonator as discussed above, 
we need to explore the Volta and patch electrostatic potentials present 
between the two surfaces. To quantify this source of noise it is worth 
introducing, in analogy to the discussion in \cite{Onofrio} for the 
parallel plane case, the equivalent voltage corresponding, at a given
distance, to the Casimir force. 
By equating Eqs. (\ref{forceELprox}) and (\ref{casimircyl}) we obtain 
an equivalent voltage as 
\begin{equation}
V_{\rm eq}= \left( \frac{\pi^2 \hbar c}{192 \epsilon_0} \right)^{1/2}
\frac{1}{d}= 13.55 \left( \frac{\mu{\rm m}}{d} \right) \; {\rm mV} .
\end{equation} 
For instance, at a targeted gap of $d=3 \mu{\rm m}$ the 
Casimir force corresponds to an equivalent voltage 
$V_{\rm eq}=4.61$ {\rm mV}, which implies that one should control 
the electrostatic stray potentials within a fraction of this
value in order to see the thermal contribution to the 
Casimir force. The capability of counterbiasing the electrostatic potential
difference at this level will require a dedicated study in a concrete 
setting once the gap distances will be reduced to few micrometers.
From this point of view it is encouraging that, as visible in
Fig. \ref{fig1}, the electrostatic force in the cylinder-plane 
configuration can exceed at large distances the corresponding 
one for the parallel plane situation. This can lead to 
larger signals in a wider range of distances and consequently to a 
more precise determination of the stray voltages to be counterbiased.  
 
The apparatus could also be adapted to the study of the validity of the 
proximity force approximation. 
As we already mentioned at the end of Section II, recent numerical
results based on the wordline approach \cite{Gies1} indicate that 
deviations from the proximity force approximation should be relevant 
starting from a ratio $d/a \simeq 0.06$ upward. 
For instance, at a distance of 5$\mu$m they should be of the order 
of 1 $\%$ for a choice of the radius of the cylinder a=100 $\mu$m. 
This suggests the use of metallic wires to intentionally look for 
the deviation from the predictions of the proximity force 
approximation \cite{Gies,Gies1}. 

\section{Conclusions}

We have discussed a novel geometry to study the Casimir force  
which more adequately addresses current issues related to the 
large distance behavior of quantum vacuum fluctuations. 
In particular, this should allow for the study of the interplay of zero-point
fluctuations with the thermal contribution due to the 
real photons present at any finite temperature. From this point of
view the theoretical discussion presented here is complementary to the 
one presented in \cite{Dalvit} for the case of eccentric cylinders, 
the latter being promising for investigating extra-gravitational 
forces due to a smaller sensitivity to long-range Casimir related effects.
An apparatus to demonstrate some of the concepts developed in this discussion 
has been built and tested, showing promising features towards 
the observation of the thermal effect. 
With an upgraded version of the apparatus, including better thermal 
stabilization, high precision actuators, and mechanical resonators with 
decreased roughness, and with a careful study of the stray
electrostatic potentials, we should be able to explore the Casimir
force in the target range centered around $3 \mu{\rm m}$ with a  
precision of a few percent. Moreover, the possible observation 
of deviations from the force predicted by the proximity force 
approximation will be important to assess the limits to 
extra-dimensional forces of gravitational origin. 

\begin{acknowledgments}

We are grateful to Astrid Lambrecht and Serge Reynaud for 
providing us with the permittivity data for different metals. 
We also thank Scott M. Middleman for experimental assistance, 
Richard L. Johnson for skillful technical support, and Holger Gies 
for  fruitful discussions and for sharing crucial information 
on worldline numerical simulations prior to publication. 
M.B.H. acknowledges support from the Dartmouth Graduate Fellowship 
program, W.J.K. acknowledges  the Gordon Hull Fellowship, and 
F.D.M. acknowledges support from Universidad de Buenos Aires, 
CONICET and ANPCyT.

\end{acknowledgments}


\begin{thebibliography}{99}

\bibitem{Milonni} P. Milonni,
{\it The Quantum Vacuum} (Academic Press, San Diego, 1994).

\bibitem{Zeldovich} Y. B. Zeldovich,
Sov. Phys. JETP \textbf{6}, 316 (1967).

\bibitem{Weinberg} S. Weinberg,
Rev. Mod. Phys. \textbf{61}, 1 (1989).

\bibitem{Carroll} S. M. Carroll, Living Rev. Rel. \textbf{4}, 1
  (2001).

\bibitem{Sahni} V. Sahni and A. A. Starobinsky, 
Int. J. Mod. Phys. D \textbf{9}, 373 (2000).  

\bibitem{Peebles} P.J.E. Peebles,
Rev. Mod. Phys. \textbf{75}, 559 (2003).

\bibitem{Casimir} H.B.G. Casimir,
Proc. K. Ned. Akad. Wet. B \textbf{51}, 793 (1948).

\bibitem{Plunien}
G. Plunien, B. M\"uller, and W. Greiner, Phys. Rep. \textbf{134}, 87
(1986).

\bibitem{Mostepanenko}
V. M. Mostepanenko and N. N. Trunov, {\it The Casimir Effect and its Applications}
(Clarendon, London, 1997)

\bibitem{Bordag} M. Bordag, {\it The Casimir Effect 50 Years
  Later} (World Scientific, Singapore, 1999).

\bibitem{Bordag1}
M. Bordag, U. Mohideen, and V. M. Mostepanenko,
Phys. Rep. \textbf{353}, 1 (2001).

\bibitem{Reynaud}
S. Reynaud, A. Lambrecht, C. Genet, M.T. Jaekel, 
C. R. Acad. Sci. Paris \textbf{IV-2}, 1287 (2001).

\bibitem{Milton}
K. A. Milton, {\it The Casimir Effect: Physical Manifestations of
  the Zero-Point Energy} (World Scientific, Singapore, 2001).

\bibitem{Milton1}
K. A. Milton, J. Phys. A: Math. Gen. \textbf{37}, R209 (2004).

\bibitem{Lamoreauxrev}
S.K. Lamoreaux, Rep. Prog. Phys. \textbf{68}, 201 (2005).

\bibitem{Sparnaay} M.J. Sparnaay, Physica \textbf{24}, 751 (1958).

\bibitem{vanBlockland} P.H.G.M. van Blokland and J.T.G. Oveerbeek,
J. Chem. Soc. Faraday Trans. I, \textbf{74}, 2637 (1978).

\bibitem{Bressi} G. Bressi, G. Carugno, R. Onofrio, and G. Ruoso,
Phys. Rev. Lett. \textbf{88}, 041804 (2002).

\bibitem{Lamoreaux} S.K. Lamoreaux, Phys. Rev. Lett. \textbf{78}, 5
  (1997).

\bibitem{Mohideen} U. Mohideen and A. Roy,
Phys. Rev. Lett. \textbf{81}, 4549 (1998);
B.W. Harris, F. Chen, and U. Mohideen,
Phys. Rev. A \textbf{62}, 052109 (2000).

\bibitem{Chan} H.B. Chan, V.A. Aksyuk, R.N. Kleiman, D.J. Bishop,
and F. Capasso, Science \textbf{291}, 1941 (2001);
H. B. Chan, V.A. Aksyuk, R.N. Kleiman, D.J. Bishop,
and F. Capasso, Phys. Rev. Lett. \textbf{87}, 211801 (2001);
D. Iannuzzi, I. Gelfand, M. Lisanti, and F. Capasso, Proc.
Nat. Ac. Sci. USA \textbf{101}, 4019 (2004).

\bibitem{Decca} R.S. Decca, D. Lopez, E. Fischbach, and D.E. Krause,
Phys. Rev. Lett. \textbf{91}, 050402 (2003); 
R.S. Decca {\it et al.}, Phys. Rev. Lett. \textbf{94}, 240401 (2005).

\bibitem{Decca1} R.S. Decca, D. Lopez, E. Fischbach,
  G. L. Klimchitskaya, D.E. Krause, and V.M. Mostepanenko,  
 Annals of Physics \textbf{318}, 37 (2005). 

\bibitem{Footnote1} In the only sphere-plane experiment exploring 
distances larger than 1$\mu$m \cite{Lamoreaux}, accuracy at the
level of $5-10\%$  at the largest explored distances was reported, see 
A. Lambrecht and S. Reynaud, Phys. Rev. Lett. \textbf{84}, 5672
(2000); S.K. Lamoreaux, Phys. Rev. Lett. \textbf{84}, 5673 (2000).
Also, common to all experiments is the issue of the precision 
obtainable in the measurement of the absolute value of the distance 
between the two objects, as further discussed in Section III.

\bibitem{Derjaguin} B. V. Derjaguin and I. I. Abrikosova,
  Sov. Phys. JETP \textbf{3}, 819 (1957);
B. V. Derjaguin, Sci. Am. \textbf{203}, 47 (1960).

\bibitem{Blocki} J. Blocki, J. Randrup, W.J. Swiatecki, and F. Tsang,
Ann. Phys. \textbf{105}, 427 (1977). 

\bibitem{Fishbach} E. Fischbach and C. L. Talmadge,
{\it The Search for Non-Newtonian Gravity}
(AIP/Springer-Verlag, New York, 1999).

\bibitem{Smythe} W.R. Smythe, {\it Static and Dynamic Electricity}
(McGraw-Hill, New York, 1968), p. 78.

\bibitem{Dalvit}
D.A.R. Dalvit, F.C. Lombardo, F.D. Mazzitelli, and R. Onofrio,
Europhys. Lett. \textbf{67}, 517 (2004).

\bibitem{Lifshitz} E.M. Lifshitz, Zh. Eksp. Teor. Fiz 
\textbf{29}, 94 (1956) [Sov. Phys. JETP \textbf{2}, 73 (1956)];
E.M. Lifshitz and L.P. Pitaevskii, Statistical Physics, Part 2 
(Butterworth-Heinemann,  Oxford, 2002).

\bibitem{Palik} Handbook of Optical Constants of Solids 
(Ed. E. D. Palik, Academic Press, London, 1985).

\bibitem{Lambrecht} A. Lambrecht and S. Reynaud, Eur. Phys. J. D \textbf{8}, 309 (2000).

\bibitem{Bostrom} M. Bostrom and B. E. Sernelius, Phys. Rev. Lett.
  \textbf{84}, 4757 (2000).

\bibitem{Lamoreaux1} S. K. Lamoreaux, Phys. Rev. Lett. \textbf{87}, 139101 (2001);
B. E. Sernelius, Phys. Rev. Lett. \textbf{87}, 139102 (2001).

\bibitem{Chen} F. Chen, G. L. Klimchitskaya, U. Mohideen, and V. M. Mostepanenko,
  Phys. Rev. Lett. \textbf{90}, 160404 (2003).

\bibitem{Geyer} B. Geyer, G. L. Klimchitskaya, and V.M. Mostepanenko, 
Phys. Rev. A \textbf{67}, 062102 (2003); {\it ibid.} \textbf{65},
062109 (2002).  

\bibitem{Esquivel} R. Esquivel, C. Villarreal, and W.L. Mochan,
Phys. Rev. A \textbf{68}, 052103 (2003).

\bibitem{Genet} C. Genet, A. Lambrecht, and S. Reynaud, Int. J. Mod.
  Phys. A \textbf{17}, 761 (2002); Phys. Rev. A \textbf{62}, 012110
  (2000). 

\bibitem{Svetovoy} V. B. Svetovoy and M. V. Lokhanin, Phys. Rev. A 
\textbf{67}, 022113 (2003).

\bibitem{Torgerson} J. R. Torgerson and S. K. Lamoreaux, 
Phys. Rev. E \textbf{70}, 047102 (2004).

\bibitem{Esquivel1} R. Esquivel and V. B. Svetovoy, Phys. Rev. A \textbf{69}, 062102 (2004).

\bibitem{Bordag2} M. Bordag, B. Geyer, G.L. Klimchitskaya, and V.M. Mostepanenko, Phys.
Rev. Lett. \textbf{85}, 503 (2000).

\bibitem{Hoye} J. S. H{\o}ye, I. Brevik, J. B. Aarseth, and 
K. A. Milton,
Phys. Rev. E \textbf{67},  056116 (2003);  I. Brevik, 
J. B. Aarseth, J. S. H{\o}ye, and K. A. Milton, 
Phys. Rev. E \textbf{71}, 056101 (2005); 
V. S. Bentsen, R. Herikstad, S. Skriudalen, I. Brevik, and 
J. S. H{\o}ye, quant-ph/0505136.

\bibitem{Footnote2} One could also consider a proximity force
  approximation in which the infinitesimal surfaces on the plane 
and on the cylinder, to which the parallel plane result is 
applied, are not the closest ones, rather the pair of 
surfaces connected by the normal to the cylinder. 
The former are separated by $d(\varphi)/\cos\varphi$. 
In this approximation there is an additional factor 
$\cos^2 \varphi$ in the proximity integral of Eq. (10).

\bibitem{Footnote3} The leading and next-to-leading
corrections are independent of the upper limit of integration
in Eq. (\ref{var-el}), because for small $d/a$ the main contribution
to the integral comes from the region $\varphi\leq 1$.

\bibitem{Stecher} F. D. Mazzitelli, M. J. Sanchez, N. N. Scoccola, 
and J. von Stecher, Phys. Rev. A \textbf{67}, 013807 (2003).

\bibitem{Gies} H. Gies, K. Langfeld, and L. Moyaerts,
J. High Energy Phys. \textbf{6}, 18 (2003).

\bibitem{Gies1} H. Gies, private communication.

\bibitem{Rugar} D. Rugar, H. J. Marmin, and P. Guethner, 
Appl. Phys. Lett. \textbf{55}, 2588 (1989).

\bibitem{Bressi2} G. Bressi, G. Carugno, A. Galvani, 
R. Onofrio, G. Ruoso, and F. Veronese, 
Class. Quantum Grav. \textbf{18}, 3943 (2001).

\bibitem{Footnote4} In the parallel plane configuration studied in 
\cite{Bressi}, parallelism was achieved through a complex system of 
relative micropositioning of the two plates, and monitoring their  
capacitance with an AC bridge. Micrometer size dust particles 
in between the plates often resulted in minimum achievable gaps 
of the order of $3-5 \mu$m, a problem already experienced in 
\cite{Sparnaay}. From this point of view a cylinder-plane 
configuration is more immune to the presence of dust particles, unless 
they are exactly located along the line of minimum distance between 
the cylinder and the plane. 

\bibitem{Onofrio} R. Onofrio and G. Carugno, Phys. Lett. A \textbf{198}, 365 (1995).
 
\end{thebibliography}
\end{document}